\documentclass[twocolumn,showpacs,preprintnumbers,amsmath,amssymb,floatfix]{revtex4}
\usepackage{graphicx}
\newcommand{\ud}{\mathrm{d}}

\begin{document}

\title{Diffusion and Current of Brownian Particles in Tilted
Piecewise Linear Potentials: Amplification and Coherence}
\author{Els Heinsalu, Risto Tammelo*, Teet \"Ord}
\affiliation{ Institute of Theoretical Physics, Tartu University,
4 T\"ahe Street, 51010 Tartu, Estonia}


\begin{abstract}

Overdamped motion of Brownian particles in tilted piecewise linear
periodic potentials is considered. Explicit algebraic expressions
for the diffusion  coefficient, current, and coherence level of
Brownian transport are derived. Their dependencies on temperature,
tilting force, and the shape of the potential are analyzed. The
necessary and sufficient conditions for the non-monotonic behavior
of the diffusion coefficient as a function of temperature are
determined. The diffusion coefficient and coherence level are
found to be extremely sensitive to the asymmetry of the potential.
It is established that at the values of the external force, for
which the enhancement of diffusion is most rapid, the level of
coherence has a wide plateau at low temperatures with the value of
the P\'{e}clet factor 2. An interpretation of the amplification of
diffusion in comparison with free thermal diffusion in terms of
probability distribution is proposed.

\end{abstract}

\vspace{1cm} \maketitle

\section{Introduction}

It is difficult to overestimate the importance of the effects
caused by Brownian motion  for  soft condensed matter physics. An
object of special attention has been Brownian motion in periodic
structures which has various applications in condensed matter
physics, chemical physics, nanotechnology,  and molecular biology
$\,$ \cite{risken,reimann,landa00,braun}. Furthermore, the idea
that adding noise to deterministic motion can give nontrivial
results has led to many important discoveries, such as stochastic
resonance \cite{gammaitoni}, resonant activation \cite{doering},
noise-induced spatial patterns \cite{garcia}, noise-induced
multistability as well as noise-induced phase transitions
\cite{landa00,broeck94,broeck97,landa96,garcia99}, ratchets
\cite{reimann}, and hypersensitive transport \cite{ginzburg}, to
name but a few of the new phenomena in this field.

The acceleration of thermal diffusion of Brownian particles in
tilted periodic potentials has been studied recently in a number
of papers \cite{constantini,lindner01,reimann01,reimann02}. An
small enhancement of diffusion induced by bias with respect to
free diffusion was established for the asymptotic limits
\cite{constantini}. The papers
\cite{reimann01,reimann02,lindner01} provided exact analytic
treatments of the effect for arbitrary temperature, tilting force,
and periodic potential. A giant amplification of diffusion up to
fourteen orders of magnitude was predicted in
Refs.~\cite{reimann01,reimann02}. Another new effect, a
non-monotonic behavior of the diffusion coefficient and coherence
level of the transport of Brownian particles as a function of
temperature was found in Ref.~\cite{lindner01}. Similar anomalies
were observed in systems with spatially periodic temperature
\cite{lindner02}. Recently it was shown that non-homogeneous
dissipation can induce enhancement and suppression of the
diffusion as a function of temperature, as well as an increase of
the coherence level of Brownian motion in tilted symmetric
periodic potential \cite{dan}. A distinctive behavior of diffusion
also occurs if one applies a sinusoidal time-periodic force to the
system, in addition to a tilted sinusoidal space-periodic force.
It was demonstrated numerically that the interplay between
frequency-locking and noise gives rise to a multi-enhancement of
the effective diffusion, and a rich behavior of the current,
including partial suppression and characteristic resonances
\cite{reguera} (see also \cite{gang}).

Usually, when addressing diffusion enhancement, tilted harmonic
spatial periodic potential is used, confer however Ref.
\cite{lindner01}. Furthermore, it is known that some features of
the ratchet transport mechanism (e.g., the current reversals) are
extremely sensitive to the shape of the potential \cite{mankin}.
In the present paper, we carry out a comprehensive study of the
dependence of diffusive and coherent motion of overdamped Brownian
particles on temperature and tilting force for various shapes of
tilted piecewise linear periodic potentials. The latter are
sufficiently simple to allow an algebraic treatment of the
relevant quantities, being at the same time physically rich enough
to provide most of the effects characteristic of tilted periodic
potentials with one minimum per period.

The paper is organized as follows. Proceeding from the general
scheme developed in Refs. \cite{reimann01,reimann02} (outlined in
Sec.~II), we derive in Sec.~III the algebraic expressions for the
diffusion coefficient, current, and P\'{e}clet factor for the case
of a general tilted piecewise linear periodic potential. The
asymptotic limits and particular cases are considered analytically
in Sec.~IV. In Sec.~V we analyze the behavior of the diffusion
coefficient and P\'{e}clet factor in the space of system
parameters, i.~e., the tilting force, temperature, and asymmetry
parameter of the potential. We also discuss the effect of the
amplification of diffusion in terms of probability distribution.
The existence of correlation between the enhancement of diffusion
and the stabilization of the coherence level of Brownian transport
is demonstrated in Sec.~VI. Our results are summarized in
Sec.~VII.

\section{general scheme}

We consider overdamped motion of Brownian particles described by
one-dimensional Langevin equation
\begin{equation} \label{1}
\eta \frac {\ud x(t)}{\ud t}=-\frac {\ud V(x)}{\ud x}+\xi(t) \, ,
\end{equation}
where $V(x) \! = \! V_{0}(x)-Fx$ and $V_{0}(x) \! = \! V_{0}(x+L)$
is the periodic potential, $F$ is the static tilting force, $\eta$
is the coefficient of viscous friction, $\xi(t)$ is the Gaussian
white noise with the mean value $\langle \xi (t) \rangle \! \! =
\! 0$ and the correlation function $\langle \xi (t) \, \xi (t')
\rangle \! = 2 \, \eta k_{B} T \, \delta(t-t')$.

As usual, the effective diffusion coefficient is defined by
\begin{equation}
D= \lim _{t \to \infty} \frac {\langle x^{2}(t) \rangle -\langle
x(t) \rangle ^{2}}{2t} \, .
\end{equation}
In the same time-scale the current is determined as
\begin{equation}
\langle \dot{x} \rangle =\lim_{t \to \infty} \frac{\langle x(t)
\rangle}{t} \, .
\end{equation}
According to Refs.~\cite{reimann01,reimann02} the diffusion
coefficient for the model (\ref{1}) is equal to ($F \! \geq 0$)
\begin{equation} \label{5}
D= \frac {D_0}{N^3} \int_{x_0}^{x_0 +L} \! I_{+}(x) \,[I_{-}(x)]^2
\frac {\ud x}{L} \, ,
\end{equation}
with
\begin{equation} \label{N}
N= \! \int_{x_0} ^{x_0+L} \! I_{-}(x) \frac {\ud x}{L} \, ,
\end{equation}
\begin{eqnarray} \label{6}
I_{\pm}(x)= \mp {D_0}^{-1} \, e^{\pm V(x)/k_{B}T} \int_{x}^{x \mp
L} \! e^{\mp V(y)/k_{B}T} \ud y \, .
\end{eqnarray}
Here $x_0$ is an arbitrary point and $D_0 \! \! = \! k_{B}T/\eta$
is the diffusion coefficient for $V_0(x) \! \! = \! 0$, i.e., for
free diffusion. The particle current is given by
\cite{reimann01,reimann02,stratonovich}
\begin{equation} \label{v}
\langle \dot{x} \rangle =N^{-1} \left( 1-e^{-LF/k_{B}T} \right) \,
.
\end{equation}

The relationship between the directed and diffusive components in
Brownian motion (the level of coherence) can be characterized by
the P\'{e}clet number $P \! e$ or by the factor of randomness $Q$
\begin{equation} \label{Qd8}
P \! e=\frac {L \langle \dot{x} \rangle}{D}= \frac {2}{Q} \, .
\end{equation}
The greater the P\'{e}clet number, the greater the coherence of
Brownian transport.

\vspace{0.7cm}
\section{analytic computations}

We proceed form the exact analytical formulas for the effective
diffusion coefficient (\ref{5}) and current (\ref{v}) of Brownian
particles, and consider a piecewise linear periodic potential with
the amplitude $A$, period $L$, and asymmetry parameter $k$ ($0 \!
\! < \!\! k \!\! < \!\! L$; the potential is symmetric if $k \!\!
= \!\! L/2$). If the tilting force exceeds the critical value
\begin{equation} \label{36}
F > F_{c}= \frac {A}{L-k} \, ,
\end{equation}
the potential is monotonically decreasing, i.e., the potential
does not have local minima.

We can take with no loss of generality $L \!\! = \! \! 1$ and
replace the relevant quantities with the corresponding
dimensionless ones: $\tilde{T}\! =\! k_{B}T A^{-1}\,$ and
$\tilde{F}\! =\! F/F_{c},\,$ hence $\tilde{F_c}\! =\! 1,\,$
$\tilde{D}\! =\! D \eta A^{-1},\,$ $\tilde{D_0} = D_0 \eta A^{-1}$
so that $\tilde{D_0} \! =\! \tilde{T}$ and $\langle
\tilde{\dot{x}}\rangle \! =\! \eta A^{-1}\, \langle \dot{x}
\rangle$. For brevity, in what follows we will omit the tilde
signs above the symbols.

Now the expressions for the diffusion coefficient, current, and
P\'{e}clet factor have the following form:
\begin{equation} \label{x}
\langle \dot{x} \rangle= \varphi_0 Z^{-1} \, ,
\end{equation}
\begin{equation} \label{Di}
D= TY Z^{-3} \, ,
\end{equation}
\begin{equation} \label{Q}
P \! e= \varphi_0 Z^2 (TY)^{-1} \, ,
\end{equation}
where
\begin{equation} \label{FI0}
\varphi_{0}=1-\exp\left({-\frac{F}{T(1-k)}}\right) \, ,
\end{equation}
\begin{equation} \label{Z0}
Z=\int_{0}^{k} \! H_{-a}(x) \, \ud x + \int_{k}^{1} \! H_{-b}(x)
\, \ud x \, ,
\end{equation}
\begin{eqnarray} \label{Y0}
Y\!\!\!&=&\!\!\!\int_{0}^{k} \! H_{+a}(x) \, [H_{-a}(x)]^{2} \,
\ud x \nonumber\\
&&+\int_{k}^{1}\! H_{+b}(x) \, [H_{-b}(x)]^{2} \,
\ud x \, .
\end{eqnarray}

Equations (\ref{Z0}) and (\ref{Y0}) contain the functions $H_{\pm
a}$ and $H_{\pm b}$ where the subscripts $a$ and $b$ associate,
correspondingly, with the limits of integration from 0 to $k$ and
from $k$ to $1$. We have
\begin{widetext}
\begin{eqnarray} \label{38a}
H_{\pm a}(x)\!=\!D_0^{-1}e^{[\pm v_{a}(x)-F(1\pm1)/2(1-k)]/T}
\biggl\{\int \limits _{x}^{k} \! e^{\mp v_{a}(y)/T} \ud y + \!
\int \limits _{k}^{1} \! e^{\mp v_{b}(y)/T} \ud y + \!
\int \limits_{1}^{x+1} \! e^{\mp v_{\tilde{a}}(y)/T} \ud y \biggr\}, \nonumber \\
H_{\pm b}(x)\!=\!D_0^{-1}e^{[\pm v_{b}(x)-F(1\pm1)/2(1-k)]/T}
\biggl\{\int \limits _{x}^{1} \! e^{\mp v_{b}(y)/T} \ud y + \!
\int \limits _{1}^{k+1} \! e^{\mp v_{\tilde{a}}(y)/T} \ud y + \!
\int \limits _{k+1}^{x+1} \! e^{\mp v_{\tilde{b}}(y)/T} \ud y
\biggr\},
\end{eqnarray}
\end{widetext}
In Eqs.~(\ref{38a}) the dimensionless potential reads as
\begin{eqnarray}
v_{a}(x)\!\!& =&\!\! a_{0}-ax \, , \quad  0 \!\leq x\!\! \leq k , \nonumber \\
v_{b}(x)\!\!&=&\!\!-b_{0}+bx \, , \quad k \!\leq x\!\! \leq 1 , \nonumber \\
v_{\tilde{a}}(x)\!\!&=&\!\!\tilde{a}_{0}-ax \, , \quad 1 \!\leq x\!\! \leq 1+k , \nonumber \\
v_{\tilde{b}}(x)\!\!&=&\!\!-\tilde{b}_{0}+bx \, , \quad 1+k\! \leq
x \!\!\leq 2  ,
\end{eqnarray}
with
\begin{eqnarray}
a\!\!&=&\!\!\frac{1-(1-F)k}{(1-k)k} \, , \quad a_{0}=1 \, ,
\quad \tilde{a}_{0}=\frac{1+k}{k} \, , \nonumber \\
b\!\!&=&\!\!\frac{1-F}{1-k} \, , \quad b_{0}=\frac{k}{1-k} \, ,
\quad \tilde{b}_{0}=\frac{1+k}{1-k} \, .
\end{eqnarray}
Performing integration in Eqs. (\ref{38a}), we obtain
\begin{eqnarray} \label{38b}
H_{\pm a}(x)\!\!&=& \!\! \frac{\varphi_{0}}{a}+g\varphi_{a} \nonumber \\
&&\times \exp\left(\frac{a[\mp 2x-k(1 \mp 1)]}{2T}\right) \, , \nonumber \\
H_{\pm b}(x)\!\!&=&\!\!-\frac{\varphi_{0}}{b}+g\varphi_{b} \nonumber \\
&&\times \exp\left(\frac{b[\pm 2(x-1)+(1-k)(1 \pm 1)]}{2T}\right) \, , \nonumber \\
\end{eqnarray}
with the notations
\begin{equation} \label{g}
g=\frac{1}{a}+ \frac{1}{b} \, ,
\end{equation}
\begin{eqnarray} \label{FI}
\varphi_{a}\!\!&=&\!\!\exp\left({\frac{1-F}{T}}\right)-1 \, , \nonumber \\
\varphi_{b}\!\!&=&\!\!
1-\exp\left({-\frac{1-(1-F)k}{T(1-k)}}\right) \, .
\end{eqnarray}

Substituting the functions $H_{\pm a, b}(x)$ from Eqs.~(\ref{38b})
into (\ref{Z0}) and (\ref{Y0}), we have after integration
\begin{equation} \label{Z}
Z= \left( \frac{k}{a} - \frac{1-k}{b} \right) \varphi_0 + T g^{2}
\varphi_{a} \varphi_{b} \, ,
\end{equation}
\begin{eqnarray} \label{Y}
Y\!\!&=&\!\!\left( \frac{k}{a^{3}} -\frac{1-k}{b^{3}}\right)
\varphi_{0}^{3}+3T\left(\frac{1}{a^{3}}+\frac{1}{b^{3}}\right)
g\varphi_{0}^{2}\varphi_{a}\varphi_{b} \nonumber\\
&& +\frac{1}{2}T g^{2}\varphi_{0}
\left[\frac{1}{a^{2}}\varphi_{a}^{2}\tilde{\varphi_{b}}-
\frac{1}{b^{2}}\varphi_{b}^{2}\tilde{\varphi_{a}}\right] \nonumber\\
&& +2g^{2}\varphi_{0}
\left[\frac{k}{a}\varphi_{a}^{2}(1-\varphi_{b})-
\frac{1-k}{b}\varphi_{b}^{2}(1+\varphi_{a})\right] \nonumber\\
&& +Tg^{3}
\left[\frac{1}{a}\varphi_{a}^{3}\varphi_{b}(1-\varphi_{b})+
\frac{1}{b}\varphi_{b}^{3}\varphi_{a}(1+\varphi_{a})\right], \nonumber\\
\end{eqnarray}
where
\begin{eqnarray}
\tilde{\varphi_{a}}\!\!&=&\!\!\exp\left({\frac{2(1-F)}{T}}\right)-1 \, , \nonumber \\
\tilde{\varphi_{b}}\!\!&=&\!\!1-\exp\left({-\frac{2[1-(1-F)k]}{T(1-k)}}\right)
\, .
\end{eqnarray}

By that we have derived the exact algebraic expressions for the
current $ \langle \dot{x} \rangle $, the diffusion coefficient
$D$, and the P\'{e}clet factor $P \! e$.

\section{asymptotic limits and particular cases}

In this Section we will examine the asymptotic limits and
essential particular cases on the basis of the analytical formulas
derived.

(i) In the absence of tilt ($F=0$), Eqs.~(\ref{FI0}) and
(\ref{FI}) reduce to
\begin{eqnarray}
\varphi_0=0 \, , \quad \varphi_a=e^{1/T}-1 \, , \quad
\varphi_b=1-e^{-1/T},
\end{eqnarray}
and from Eqs.~(\ref{Di}), (\ref{Z}), and (\ref{Y}) one obtains
\begin{equation} \label{D0}
D= \frac{1}{2T \, [\cosh(1/T)-1]} \, .
\end{equation}
This expression is as a special case of the general formula of the
diffusion coefficient for arbitrary unbiased periodic potential
\cite{lifson}. It is to be noticed, that for $F=0$ the coefficient
of diffusion becomes independent of the asymmetry parameter $k$.

(ii) In the high temperature limit, one can  take into account
only the first order terms in the expansions of the exponents in
Eqs.~(\ref{FI0}) and (\ref{FI}). Then
\begin{equation}
\varphi_{0} \approx \frac{F}{T(1-k)} , \;\;
\varphi_{a}\approx\frac{1-F}{T}, \; \;
\varphi_{b}\approx\frac{1-(1-F)k}{T(1-k)}
\end{equation}
and
\begin{equation}
H_{\pm a,b}=T^{-1} \, , \quad Z=T^{-1} \, , \quad Y=T^{-3}.
\end{equation}
The current, diffusion coefficient, and P\'{e}clet factor now
become
\begin{equation} \label{7}
\langle \dot{x} \rangle= \frac{F}{1-k} \, ,
\end{equation}
\begin{equation} \label{8}
D=T \, ,
\end{equation}
\begin{equation} \label{9}
P \! e=\frac{F}{T(1-k)} \, .
\end{equation}

(iii) Under the conditions $F\! \gg \! 1$ and $F\!/T \! \gg \! 1$,
it is valid that
\begin{equation}
\varphi_{0}\approx-\varphi_{a}\approx\varphi_{b}\approx 1 \, ,
\quad a\approx-b\approx \frac{F}{1-k}\;,
\end{equation}
and
\begin{equation}
H_{\pm a,b}=\frac{1-k}{F} \, , \quad Z \! = \! \frac{1-k}{F} \,
,\quad Y \!  = \! \left(\frac{1-k}{F}\right)^{3}.
\end{equation}
As a result, the expressions for $D$, $\langle \dot{x} \rangle$,
and $P \! e$ coincide with Eqs.~(\ref{7})-(\ref{9}). Thus, at high
temperatures and at large values of tilting force, the transport
properties of Brownian particles are the same.

(iv) If $F \! < \! 1$ and $(1-F)/T \! \gg \! 1$, we have the
following asymptotic limits:
\begin{eqnarray}
\varphi_{b} \!\!& \approx & \!\!\tilde{\varphi}_{b} \approx 1 \, , \nonumber \\
\varphi_{a} \!\!& \approx & \!\!e^{(1-F)/T}\gg 1 \, , \nonumber \\
\tilde{\varphi}_{a}\!\! & \approx &
\!\!e^{2(1-F)/T}=\varphi_{a}^{2} \, .
\end{eqnarray}
Then   Eqs.~(\ref{Z0}), (\ref{Y0}), and (\ref{38b}) yield
\begin{equation}
Z=T g^{2} e^{(1-F)/T} \, ,
\end{equation}
\begin{eqnarray}
Y\!\!&=&\!\!\frac {1}{2} T g^{3} e^{2(1-F)/T}[(a^{-1} - b^{-1}) \varphi_0 \nonumber\\
&& + 2( a^{-1} e^{-F/(1-k)T} + b^{-1})] \, ,
\end{eqnarray}
and
\begin{equation} \label{100}
\langle \dot{x} \rangle= \frac{\varphi_0}{T g^{2} e^{(1-F)/T}} \, , \\
\end{equation}
\begin{equation} \label{101}
D= \frac{2-\varphi_0}{2T g^{2} e^{(1-F)/T} } \, , \\
\end{equation}
\begin{equation} \label{102}
P \! e=\frac{2\varphi_0}{2-\varphi_0} = 2 \tanh \frac{F}{2T(1-k)}
\end{equation}
(cf.~also~\cite{lindner01}). If, additionally, the condition $F \!
/T(\!1 \!\! - \! k) \!\!\! \gg \!\! 1$ is fulfilled, it is valid
that $e^{-F/T(1 \!-\! k)} \!\! \approx \!\! 0$ and $\varphi_0 \!\!
\approx \!\! 1$. Consequently, in the present case we have
$2D\!\!=\!\langle \dot{x} \rangle$ and $P \! e\!\!=\!2$. This
indicates that an extremely exact stabilization of the level of
coherence of Brownian transport occurs in this region of
parameters.

(v) At the critical tilt ($F\!\!=\!1$), it is valid that
\begin{eqnarray} \label{H4}
H_{\pm a}(x)\!\!&=& \!\!\frac{\varphi_0}{a}+ \frac{1-k}{T}
\exp\left(\frac{a[\mp 2x-k(1 \mp 1)]}{2T}\right) \, , \nonumber \\
H_{\pm b}(x)\!\!&=& \!\!\frac{\varphi_0}{a}+ \frac{1-k}{T}
\exp\left(-\frac{ak}{T}\right) \nonumber \\
&&+\frac{\varphi_0 [\pm 2(x-1)+(1-k)(1 \pm 1)]}{2T} \, ,
\end{eqnarray}
whereas in the low-temperature limit Eqs.~(\ref{x})-(\ref{Y0}) and
(\ref{H4}) yield
\begin{equation}
\langle \dot{x} \rangle= \frac{2T}{(1-k)^{2}} \, ,
\end{equation}
\begin{equation}
D= \frac{2T}{3(1-k)^{2}} \, ,
\end{equation}
\begin{equation}
P\!e=3 \, .
\end{equation}
We observe that, for $F\!\!=\!1$ the P\'{e}clet factor is constant
and depends neither on the temperature nor on the asymmetry
parameter.

\section{non-monotonic behavior of diffusion and coherence}
The expression of the diffusion coefficient as a function of $F$
given by Eqs.~(\ref{Di}), (\ref{Z}), and (\ref{Y}) reveals a
qualitatively similar behavior to that found in Refs.~[12,13],
exhibiting a resonant-like maximum if the temperature is
sufficiently low. This effect is strongly influenced by the shape
of the periodic potential, as illustrated in Fig.~1$\,$
%
\begin{figure}[ht]
\begin{center}
\includegraphics[width=0.8\linewidth, height=0.5\linewidth]%
{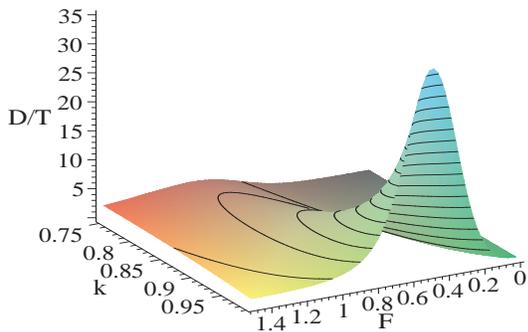}%
\caption{The plot of the diffusion coefficient \textit{vs} the
tilting force and the potential asymmetry parameter at fixed $T \!
=\! 0.1$.}
\end{center}
\end{figure}
\begin{figure}
\begin{center}
\includegraphics[width=0.8\linewidth, height=0.5\linewidth]%
{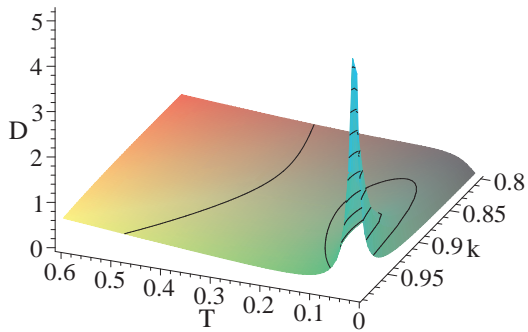}%
\caption{The plot of the diffusion coefficient \textit{vs} the
temperature and the potential asymmetry parameter at fixed $F\!
=\! 0.95$.}
\end{center}
\end{figure}
This plot shows that for positive bias ($F\!>\!0$), the increase
of the value of $k$ favors the amplification of diffusion compared
to free thermal diffusion.

Next we consider the dependence of diffusion on temperature.
Figure 2 displays the typical behavior of the diffusion
coefficient as a function of temperature and asymmetry parameter
for $F\!\!=\!0.95$. The non-monotonic behavior occurs if the value
of tilting force lies within a certain range $F_1\! <\! F\! <\!
F_2$ where $F_{1,2}$ depend on $k$. One can see that the maximum
of $D(T)$ becomes rapidly narrower and higher as $k$ approaches
unity. There exist the limiting values $k_E \! \approx 0.8285$ and
$F_0 \! \approx 1.1292$: for $k\!<\! k_E $ or $F\!>\!F_0 $ the
diffusion coefficient is a monotonic function of temperature for
all values of bias and asymmetry coefficient. Therefore, the
($k,F$)-space is divided into two domains where the analytical
properties of the diffusion coefficient as a function of
temperature are qualitatively different.

The curves of the P\'{e}clet factor \textit{versus} temperature
are depicted in Fig.~3. The function $P\!e(T)$ passes through a
maximum (curves~1-5) for $F\!\!\!<\!\!\!F_c$, which is also
present slightly above the critical tilt (curve~6). With a further
increase of $F$, the maximum of $P\!e(T)$ disappears.
\begin{figure}
\begin{center}
\includegraphics[width=0.8\linewidth, height=0.6\linewidth]%
{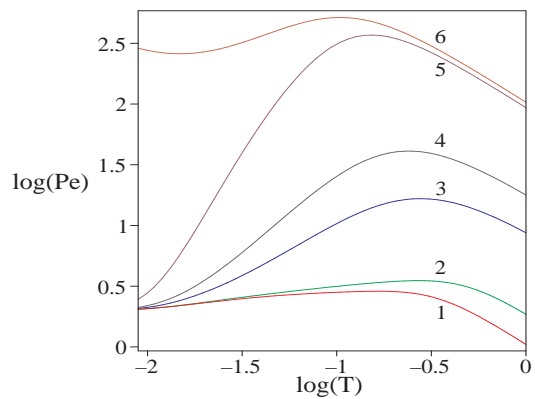}%
\caption{The plot of the P\'{e}clet factor $\log_{10}(Pe)$
\textit{versus} temperature $\log_{10}(T)$ for various values of
the potential asymmetry coefficient at $F\! =\! 0.95$ (curves 1-5)
and $F\! =\! 1.05$ (curve 6). Curve 1: $k\! =\! 0.1$, curve 2:
$k\! =\! 0.5$, curve 3: $k\! =\! 0.9$, curve 4: $k\! =\! 0.95$,
curves 5,
6: $k\! =\! 0.99$.}%
\end{center}
\end{figure}
\begin{figure}
\begin{center}
\includegraphics[width=0.8\linewidth, height=0.6\linewidth]%
{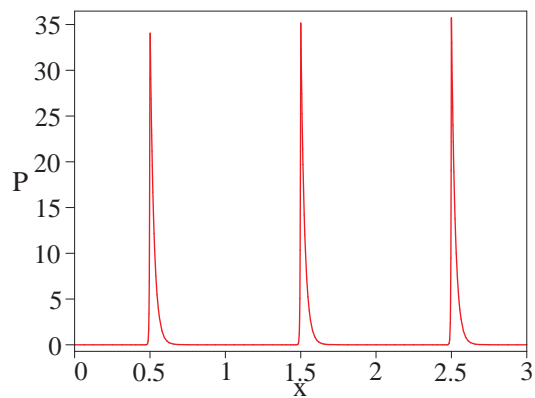}%
\caption{The plot of the probability density $P(x)$ at
$T\!=\!0.01$, $k=0.5$ and $F\!=\!0.8$.}
\end{center}
\end{figure}
As seen in Fig.~3, the optimal level of Brownian transport,
determined by the maximal value of the P\'{e}clet number, is
sensitive to the shape of periodic potential: at $F\!\!<\!\!F_c$
the optimal level of transport rises with an increase in $k$. At
the same time, the larger values of $k$ make a minimum of $D(T)$,
which follows a maximum of $D(T)$ at a higher value of
temperature, deeper (the effect can be anticipated in Fig.~2). In
this sense the situation is analogous to the results of
Ref.~\cite{dan}, where the enhancement of the coherence of
Brownian motion in a certain region of temperature due to
frictional inhomogeneity is associated with the suppression of
diffusion by the same factor.

Now we discuss the amplification of diffusion by bias in terms of
probability distribution. The stationary probability density for
the coordinate of a Brownian particle \cite{risken} normalized
over one period can be written as
\begin{equation} \label{pe77}
P(x)\!=\!\frac {1}{N} \, e^{-V(x)/k_{B}T} \! \int_{x}^{x+L}
\! e^{V(y)/k_{B}T} \, \ud y \, .
\end{equation}
Figures 4-6 represent (in terms of the dimensionless parameters)
the probability distributions characteristic to various diffusion
levels depending on the tilting force.
\begin{figure}
\begin{center}
\includegraphics[width=0.8\linewidth, height=0.6\linewidth]%
{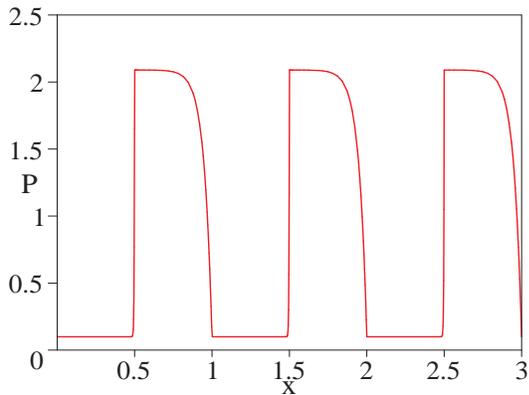}%
\caption{The plot of the probability density $P(x)$ at
$T\!=\!0.01$, $k=0.5$ and $F\!=\!1.1$.}
\end{center}
\end{figure}
\begin{figure}
\begin{center}
\includegraphics[width=0.8\linewidth, height=0.6\linewidth]%
{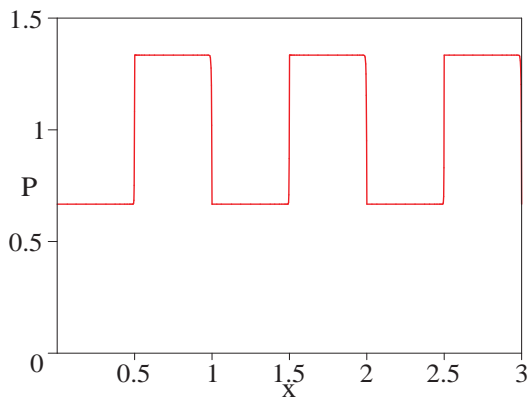}%
\caption{The plot of the probability density $P(x)$ at $T\! =\!
0.01$, $k\! =\! 0.5$ and $F\! =\! 3$.}
\end{center}
\end{figure}

Figure 4 illustrates the situation, where particles are mainly
localized around the minima of the potential and transport is
strongly suppressed. Diffusion is essentially weaker in comparison
with free diffusion: $D/T \!\! \sim \!\! 10^{-6}$. The probability
distribution shown in Fig.~5 corresponds to the case where the
diffusion is approximately maximal ($D/T \!\! =\! 3.5$) for the
chosen values of temperature and asymmetry coefficient. In this
case the regions with a large probability are separated by the
domains where the probability is much smaller, however, large
enough to allow the entrance of a sufficient number of particles
into these domains. As a result, a channel of hopping-like
transport is formed, leading to the enhancement of diffusion with
respect to free diffusion. The further increase of the tilting
force $F$ makes the probability distribution still more
homogeneous, as seen in Fig.~6, and the diffusion approaches the
free diffusion limit ($D\!/T\!\!=\!1.3$ for the values of the
parameters used in Fig.~6).

Consequently, the amplified diffusion in the tilted periodic
potential is characterized by the specific inhomogeneous
probability distribution with spatially alternating domains of
high and low probability. The occurrence of a maximum in the
dependence $D(T)$ can be understood in a similar way.

\section{correlation between the enhancement of diffusion and stabilization of coherence}

With regard to the simultaneous enhancement of diffusion and
current, caused by the force $F$, with respect to an untilted
system, the relation between $D$ and $\langle \dot{x} \rangle$ is
of interest. One can expect that such a relationship reflects some
intrinsic features of the mutual influence between diffusion and
current driven by the tilt merely at lower temperatures, when the
initial suppression of both components of Brownian transport by
periodic potential is stronger.
\begin{figure}
\begin{center}
\includegraphics[width=0.75\linewidth, height=0.5\linewidth]%
{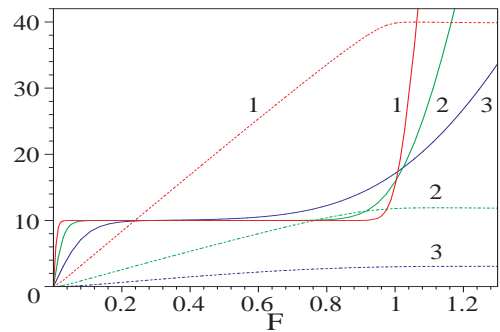}%
\caption{The comparison of the dependencies of the P\'{e}clet
factor and diffusion coefficient on the tilting force for various
temperatures at fixed $k\! =\! 0.5$. Solid lines: $5\! \times \!
P\! e$ \textit{vs} $F$, dashed lines: $\log_{10}[D(F)/D(0)]$
\textit{vs}
$F$. Curves~1: $T\! =\! 0.01$, curves~2: $T\! =\!  0.03$, curves~3: $T\! =\! 0.09.$}%
\end{center}
\end{figure}
\begin{figure}
\begin{center}
\includegraphics[width=0.75\linewidth, height=0.5\linewidth]%
{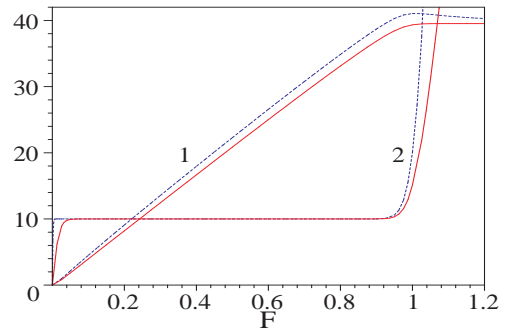}%
\caption{The comparison of the dependencies of the P\'{e}clet
factor and diffusion coefficient on the tilting force for $k\! =\!
0.1$ (solid lines) and $k\! =\! 0.9$ (dashed lines) at fixed $T\!
=\! 0.01$. Curves~1: $\log_{10}[D(F)/D(0)]$ \textit{vs} $F$,
curves~2: $5\! \times \! Pe$
\textit{vs} $F$.}%
\end{center}
\end{figure}

The comparative plot of $D$ and $P\!e$ \textit{versus} $F$ is
presented in Figs.~7 and 8. One can see that the function
$P\!e(F)$ has a point of inflection which turns into a wide
plateau at lower temperatures. For values of $F$, from zero up to
the end of the plateau, the behavior of $P\!e(F)$ is described
with great accuracy by Eq.~(\ref{102}). Note especially that the
domain where $P\!e=2$ coincides with the domain where the increase
of diffusion coefficient as a function of $F$ is most rapid and
follows with a good approximation the law
$D=c^{\alpha_{1}+\alpha_{2} F}$, where $c$ and $\alpha_{1,2}$
depend on $T$ and $k$. The location of the end of this region at
larger values of $F$ is quite insensitive to the shape of the
periodic potential, as seen in Fig.~8, and is located
approximately at $F_c$.

\section{conclusion}

In the present work we have addressed the problem of overdamped
motion of Brownian particles on tilted piecewise linear potentials
in the presence of white thermal noise. We have derived the exact
algebraic expressions for the diffusion coefficient, particle
current, and factor of randomness, which are valid for an
arbitrary simple sawtooth potential, tilting force, and strength
of the thermal noise. On the basis of these formulas we have
discussed the characteristic limiting cases and asymptotic
behavior.

Acceleration of diffusion turns out to be very sensitive to the
shape of the piecewise linear potential. It is shown that large
values of the asymmetry parameter $k$ favor the amplification of
diffusion by means of biased potential and temperature in
comparison with free thermal diffusion. This can be understood as
a result of the formation of probability distribution with
spatially alternating regions of specifically balanced high and
low probability. The necessary and sufficient conditions for the
non-monotonic behavior of the diffusion coefficient as a function
of temperature are established. On the basis of these results,
among the other applications, one can expect the significant
increase of diffusive transport, for example, in the superionic
conductors \cite{superionic1,superionic} in a strong external
electric field.

We have demonstrated that the P\'{e}clet factor as a function of
the tilting force has a plateau at low temperatures which
terminates with a steep rise at the critical value of the tilting
force. As temperature grows, the plateau  gradually reduces until
disappears and the P\'{e}clet factor becomes monotonically
increasing. The domain, where the P\'{e}clet factor exhibits the
plateau, coincides with the domain where the enhancement of
diffusion coefficient is maximal. Consequently, in the region of
parameters where substantial acceleration of diffusion occurs, the
current and diffusion are exactly synchronized. Note that the
stabilization of the coherence level of Brownian transport takes
place at the value of the P\'{e}clet factor $P\!e=2$ which is also
characteristic for Poisson processes \cite{risken}, such as the
Poisson enzymes in kinesin kinetics
\cite{svoboda,visscher,schnitzer}.

Finally, it is remarkable that, as a result of the interplay of
periodic potential, bias and white noise, there exists an exact
correlation between the acceleration of diffusion, induced by the
tilting force, and the stabilization of the coherence level of
Brownian motion. It seems that this phenomenon is quite universal
and manifests itself for an arbitrary periodic potential in
situations where initially strongly suppressed transport is
enhanced by bias, which generates significant amplification of
diffusion in comparison with free diffusion.

\vspace{0.3cm} \textbf{Acknowledgement}

The authors are grateful to Romi Mankin and Marco Patriarca for
valuable discussions, and acknowledge support by Estonian Science
Foundation through Grant No. 5662.


\begin{references}
\begin{frenchspacing}

\bibitem[*]{byline} E-mail address: {\tt tammelo@ut.ee}

\bibitem{risken} H.~Risken, \textit{The Fokker-Planck Equation}
(Springer-Verlag, Berlin, 1996).

\bibitem{reimann} P.~Reimann, Phys. Rep. \textbf{361}, 57 (2002).

\bibitem{landa00} P.~S.~Landa and P.~V.~E.~McClintock, Phys. Rep. \textbf{323}, 1
(2000).

\bibitem{braun}  O.~M.~Braun and Yu.~S.~Kivshar, Phys. Rep. \textbf{306}, 1 (1998).

\bibitem{gammaitoni} L.~Gammaitoni, P.~H\"anggi, P.~Jung, and F.~Marchesoni, Rev. Mod. Phys. \textbf{70}, 223 (1998).

\bibitem{doering} C.~R.~Doering and J.~C.~Gadoua, Phys. Rev. Lett. \textbf{69}, 2318 (1992).

\bibitem{garcia} J.~Garcia-Ojalvo, A.~Hernandez-Machado, and J.~M.~Sancho, Phys. Rev. Lett.
\textbf{71}, 1542 (1993).

\bibitem{broeck94} C.~Van den Broeck, J.~M.~R.~Parrondo, and R.~Toral, Phys. Rev. Lett. \textbf{73}, 3395
(1994).

\bibitem{broeck97} C.~Van den Broeck, J.~M.~R.~Parrondo, R.~Toral, and
R.~Kawai, Phys. Rev. E \textbf{55}, 4084 (1997).

\bibitem{landa96} P.~S.~Landa,  \textit{Nonlinear Oscillations and waves in dynamical
systems} (Kluwer, Dordrecht, 1996).

\bibitem{garcia99} J.~Garcia-Ojalvo and J.~M.~Sancho, \textit{Noise in spatially extended systems}
(Springer, Berlin, 1999).

\bibitem{ginzburg} S.~L.~Ginzburg and M.~A.~Pustovoit, Phys. Rev.
Lett. \textbf{80}, 4840 (1998).

\bibitem{constantini} G.~Constantini and F.~Marchesoni, {Europhys. Lett.} \textbf{48}, 491
(1999).

\bibitem{reimann01} P.~Reimann, C.~Van den Broeck, H.~Linke, P.~H\"anggi,
J.~M.~Rubi, and A.~P\'erez-Madrid, Phys. Rev. Lett. \textbf{87},
010602 (2001).

\bibitem{reimann02} P.~Reimann, C.~Van den Broeck, H.~Linke, P.~H\"anggi,
J.~M.~Rubi, and A.~P\'erez-Madrid, Phys. Rev. E \textbf{65},
031104 (2002).

\bibitem{lindner01} B.~Lindner, M.~Kostur, and L.~Schimansky-Geier, Fluct. Noise Lett.
\textbf{1}, R25 (2001).

\bibitem{lindner02} B.~Lindner and L.~Schimansky-Geier, Phys. Rev. Lett. \textbf{89},
230602 (2002).

\bibitem{dan} D.~Dan and A.~M.~Jayannavar, Phys. Rev. E \textbf{66}, 041106 (2002).

\bibitem{reguera} D.~Reguera, P.~Reimann, P.~H\"anggi, and J.~M.~Rubi, Europhys. Lett. \textbf{57}, 644 (2002).

\bibitem{gang} H.~Gang, A.~Daffertshofer and H.~Haken, Phys. Rev. Lett. \textbf{76}, 4874 (1996).

\bibitem{mankin}
R.~Mankin, A.~Ainsaar, and E.~Reiter, Phys.Rev. E {\bf61} 6359
(2000); R.~Mankin, A.~Ainsaar, A.~Haljas, and E.~Reiter,
{\it{ibid.}} {\bf63} 041110 (2001); R.~Mankin, R.~Tammelo, and
D.~Martila, {\it{ibid.}} {\bf64}, 051114 (2001); R.~Tammelo,
R.~Mankin, and D.~Martila, {\it{ibid.}} {\bf66}, 051101 (2002);
R.~Mankin, A.~Haljas, R.~Tammelo, and D.~Martila, {\it{ibid.}}
{\bf68}, 011105 (2003).

\bibitem{stratonovich} R.~L.~Stratonovich, Radiotekh. Elektron. \textbf{3}, 497 (1958).

\bibitem{lifson} S.~Lifson and J.~L.~Jackson, J. Chem. Phys. \textbf{36}, 2410 (1962).

\bibitem{superionic1} M.~B.~Salamon (Ed.),  \textit{The Physics of Superionic Conductors}
(Springer-Verlag, Berlin, Heidelberg, 1979).

\bibitem{superionic} J.~W.~Perram (Ed.),  \textit{The Physics of Superionic Conductors
and Electrode Materials} (Plenum Press, New York, 1983).

\bibitem{svoboda} K.~Svoboda, P.~P.~Mitra, and S.~M.~Block, Proc. Natl. Acad. Sci. USA
\textbf{91}, 11782 (1994).

\bibitem{schnitzer} M.~J.~Schnitzer and S.~M.~Block,
Nature \textbf{388}, 386 (1997).

\bibitem{visscher} K.~Visscher, M.~J.~Schnitzer, and S.~M.~Block,
Nature \textbf{400}, 184 (1999).

\end{frenchspacing}
\end{references}
\end{document}